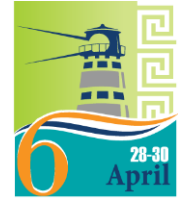

# Hospitality Students ' Perceptions towards Working in Hotels: a case study of the faculty of tourism and hotels in Alexandria University

**Sayed S. El-Houshy**

*Demonstrator, Department of Hotel Studies, Faculty of Tourism & Hotels, University of Alexandria, Egypt*

## Abstract

The tourism and hospitality industry worldwide has been confronted with the problem of attracting and retaining quality employees. If today's students are to become the effective practitioners of tomorrow, it is fundamental to understand their perceptions of tourism employment. Therefore, this research aims at investigating the perceptions of hospitality students at the Faculty of Tourism in Alexandria University towards the industry as a career choice. A self-administered questionnaire was developed to rate the importance of 20 factors in influencing career choice, and the extent to which hospitality as a career offers these factors. From the results, it is clear that students generally do not believe that the hospitality career will offer them the factors they found important. However, most of respondents (70.6%) indicated that they would work in the industry after graduation. Finally, a set of specific remedial actions that hospitality stakeholders could initiate to improve the perceptions of hospitality career are discussed.

## Introduction

Tourism is a labor-intensive service industry and its success is dependent on the availability of good-quality personnel to deliver, operate, and manage the tourist product (Amoah & Baum, 1997). The growing importance of this sector is leading to issues surrounding the number of trained personnel available to fill the growing number of positions that are becoming available in the industry (Richardson & Butler, 2011). At the same time, the industry worldwide has been confronted with the problem of attracting and retaining high quality employees, which has led to a shortage of skilled personnel to staff the large number of tourism and hospitality businesses (Powell, 1999; Hinkin & Tracey, 2000; Kusluvan & Kusluvan, 2000; Ferris *et al.,* 2002). This problem is complex with many different contributing factors, and negative disposition toward the industry is one of them (Penny & Frances, 2011).

Riley *et al.* (2002) claimed that the image of a particular industry will have a major effect on potential recruits perceptions of the industry, which will impact on the quality and quantity of future staff. The tourism and hospitality industry has a negative image in the eyes of potential recruits (Getz, 1994; Kusluvan & Kusluvan, 2000; Aksu & Koksal, 2005). While, a small number of tourism graduates pursue the tourism industry upon graduation (McKercher *et al.,* 1995; King *et al.,* 2003). A smaller percentage remained after 5 years of working in the industry (McKercher *et al.*, 1995). And about 50% of the graduates who entered the tourism industry upon graduation quit their first job and found employment outside the industry after 2 years of working in the industry (Zou *et al.,* 2002; Lu & Adler, 2009). This results in high staff turnover and waste of trained and experienced personnel (Pavesic & Brymer, 1990; Doherty *et al.,* 2001; Jenkins, 2001). This turnover becomes a wastage for all parties involved including the **1)** government that has invested money in tertiary education, **2)** the students whom spent years studying tourism courses, **3)** the tourists receiving the services from the employee, and **4)** the national economy from receiving any revenue from the repeat tourists (Ahmad *et al.,* 2009).

Students who are well-educated, well trained, and skilled continue to be a highly desirable source of talent in today's hospitality and tourism job market (Canny, 2002; Ng & Burke, 2006). However, there are studies indicating that the proportion of workers in the tourism and hospitality industry who have tertiary qualifications is much lower than most other industry sectors (Purcell & Quinn, 1996; Australian Bureau of Statistics, 2006). The importance of examining students' attitudes toward the industry lies in the fact that having a skilled, enthusiastic, and committed workforce is vital to the success of firms in the hospitality industry (Kusluvan & Kusluvan, 2000; Richardson, 2010; Penny& Frances, 2011). Knowing the values and expectations of these students could also allow hospitality programs and faculty "to guide them into right employment settings and this will ensue person–organizational fit" (Aycan & Fikret-Pasa, 2003, p.142). Also, it is important for the university to obtain information due to forecasting the number of graduates, who intend to enter the industry, especially after committing resources in developing future graduates, (Ahmad *et al.,* 2009).

Roney and Öztin (2007) claimed that if today's students are to become the effective practitioners of tomorrow, it is fundamental to understand their perceptions of tourism employment. They are the potential supply of labor in the market, and having positive attitudes will more likely lead to greater attraction and retention of these graduates in the industry (Penny & Frances, 2011). However, when looking at the overall perceptions and attitudes of tourism and hospitality management students there is relatively little evidence that research has been conducted in this area (Kusluvan & Kusluvan, 2000; Jenkins, 2001; Barron *et al.,* 2007; Roney & Öztin, 2007).

*Therefore,* this research aims at investigating the perceptions and attitudes of hospitality students who are currently enrolled in the faculty of tourism towards the hospitality industry in Egypt and its career prospects. To date, there has been no published work on this subject despite its importance. Using Alexandria University as a case study, this paper will fill this gap and provide a base for further research in this area.

**Literature Review**

- **Career Challenges in the Hospitality Industry**

The tourism industry is facing more than ever greater challenges in attracting skilled and motivated staff than the emerging sectors in the economy (Kelley-Patterson & George, 2001). Roney & Öztin (2007) claimed that although the development of the tourism industry can create new employment opportunities, it is often criticized for providing primarily low-skilled and low-paying jobs. This industry has long been associated with poor image and lack of understanding on the opportunities offered (Ahmad et al., 2009). Traditionally there has been a career-for-life philosophy adopted by workers, whereby workers will spend their entire working life working in one industry, and, in many cases, one organization (Ayres, 2006). This philosophy has, in recent times been replaced by a more uncertain career structure; with employees frequently changing employers within their industry and many also pursuing work in different industries (Inkson et al., 1999).

The vast majority of the literature reported that students have negative expectations of their future career and career prospects (Penny & Frances, 2011). Kusluvan and Kusluvan (2000) found that about 70% of respondents believed that promotion opportunities were not satisfactory and it was very difficult to find a stable job due to the seasonality factor (87%). Similarly, Richardson's study (2008) also revealed that more than half of the respondents (52.4%) claimed that the pay for most tourism jobs and the level of fringe benefits offered were low (50%). Respondents also perceived that promotions were not handled fairly and equitably (53.5%) and there was a lack of a clear career path offered by the industry (40%). Also, stable employment in the industry was difficult due to the seasonality factor (37.2%). It has been argued that this poor image is impeding the recruitment of quality staff as many potential employees are anxious about the working conditions in the industry caused by this negative portrayal (Kusluvan and Kusluvan, 2000; Aksu and Koksal, 2005).

- **Perceptions and Attitudes towards the Hospitality Career**

Ajzen (1993) as cited in Ahmad *et al.* (2009) described attitudes as "an individual's disposition to react with a certain degree of favorableness or unfavorable to an object, behavior, person or event to any other discernible aspect of the individual's world". Many researchers have studied theories on student attitudes, expectations and career choice from many different viewpoints. Domonte and Vaden (1987) have ranked the following influences that are considered to have the greatest influence on a graduate's decision to work in the tourism and hospitality industry: (1) interesting work; (2) advancement potential; (3) secure future; (4) good salary; (5) opportunity for service to society; and (6) social prestige. This concurs with McCleary and Weaver (1988) and Sciarini (1997), who agreed that type of work and advancement opportunities are the most important factors in a graduate's decision to accept a position, with most graduates expecting promotions within 2 years of graduation.

Ross (1994) found a high level of interest in management positions in the tourism industry. This is in contrast to Getz (1994) who concluded that perceptions towards a potential career in tourism had become much more negative over a period of 14 years. Purcell and Quinn (1995) surveyed 704 former tourism students and discovered that graduates complained of having little opportunity to develop their managerial skills. Several researchers have also studied the perceptions of undergraduate tourism and hospitality management students. Pavesic and Brymer (1990) found that a substantial number of graduates leave the industry owing to poor working conditions, and Barron and Maxwell (1993) examined the perceptions of new and continuing students at Scottish higher education institutions. They found that in general the new students had positive images of the industry, whereas the students with supervised work experience were much less positive in their views. Most of these studies have argued that direct experience in the tourism and hospitality industry may cause students to hold negative views of the industry.

Casado (1992) and Sciarini (1997) found that on graduation most graduates believed they were qualified enough to work as an assistant manager and were looking for a position that was of a managerial level rather than an hourly operational position. This is in contrast to Rimmington's (1999, p. 187) claim that "all graduates should be prepared to work in kitchens and restaurants to acquire practical skills. They should recognize that with that kind of grounding they will be in an excellent position to reach a senior level". At the end of their survey of high school students in Arizona, Cothran and Combrink (1999) stated that although minority students often had less knowledge about hospitality jobs, they had more interest in them. A study, conducted by Kusluvan and Kusluvan (2000), of four-year tourism and hotel management students, in seven different schools in Turkey, reported negative perceptions towards different dimensions of working in tourism. Kozak and Kızılırmak (2001) carried out a similar survey among the undergraduate tourism students in three different vocational schools in Turkey. Like Barron and Maxwell, they too indicated that work experience as a trainee in the industry affected their perceptions in a negative way.

In a study conducted by Richardson (2008) possibly the most alarming finding to come out of this study is that more than 50% of respondents are already contemplating careers outside the industry. And this concurs with the results of Richardson & Butler (2011), who explored Malaysian undergraduate tourism and hospitality students' views of the industry as a career choice, from the results, it is clear that students generally do not believe that a career in tourism and hospitality will offer them the factors that they find important. This is in contrast to Lu & Adler (2009), who showed that a majority of the undergraduate students were interested in pursuing a career in the hospitality and tourism industry. In other cases, Roney & Öztin (2007) showed that, overall, the respondents' perceptions are neither favorable nor unfavorable towards the hospitality career. Based on the aforesaid, it seems clear that both positive and negative perceptions towards the hospitality and tourism industry as a career choice were documented by many researchers in different countries and this uncovers the need for investigating the case of hospitality students in Egypt and their perceptions towards the industry.

**Research Questions**

Based on the previous literature review the following research questions were formulated to reflect the study's primary purpose:

**RQ1:** Do hospitality students intend to pursue their career in the industry after graduation?
**RQ2:** What factors students found important when considering a career?
**RQ3:** And how well they think the tourism and hospitality industry offers these factors?
**RQ4:** Have contingency variables (gender, year of study, work experience, and study major willingness) been correlated to students' current perceptions towards the industry.

**Research Methodology**

In order to precisely and fruitfully decide upon the research methodology, research objectives should be identified. This research has four main objectives which are;

1. Investigating students' current perceptions towards the hospitality industry as a career choice.

2. Understanding the most important factors to be considered when selecting a career from students' view point.

3. Discussing the relationship between gender, year of study, work experience, willingness to study tourism and students perceptions towards working in hotels.

4. Providing a set of specific remedial actions that could be initiated by hospitality stakeholders to improve the image of the industry as a career choice.

- **Sampling Procedures**

In order to achieve the objectives above, a sample of hospitality students enrolled in hotel studies department was selected. In view of the practical difficulties of employing a systematic random sampling technique in choosing the targeted sample, a non-probability convenience sampling method was used. Convenience sampling is defined as selecting the items from the population based on accessibility and ease of selection (Groebner *et al.,* 2005). The targeted population in this research includes all under and postgraduate students who study hotel management (2nd year, 3rd year, 4th year and postgraduate students) at the faculty of tourism and hotels, University of Alexandria. 1$^{st}$ year students were excluded due to the fact that specialization in the faculty begins in the second year. Due to the limitation of time for this research it was impossible to include all faculties of tourism in Egypt. Therefore Alexandria University was chosen as a case study that will represent a base for further research on this area in the future.

The population frame was adopted from the faculty students' affaires and postgraduate departments. There are an estimated 203 students enrolled in hotel studies department (Table 1 provides an overview of the survey sample by grade).

*Table 1: the survey sample and population*

| Grade | Population | Sample | Percent (%) |
|---|---|---|---|
| • Sophomore | 41 | 22 | 25.9 % |
| • Junior | 71 | 27 | 31.8 % |
| • Senior | 61 | 16 | 18.8% |
| • Postgraduate | 30 | 20 | 23.5% |
| *Total* | *203* | *85* | *100%* |

\* Population includes all students

- **Data Collection**

A self-administrated questionnaire was used to collect data from respondents. A total of 97 questionnaires were distributed in the classrooms during March 2014. Participation was totally on a voluntary basis. All the questionnaires were returned and 85 of them were found to be valid and suitable for analysis where no missing values and no inconsistent answers were found. This has yielded a high response rate of (87.6%) because the survey was conducted in a class setting.

The questionnaire was composed of two sections. The first section comprised of eight questions designed to elicit the characteristics of the respondents. The questions were adapted from a survey conducted by Roney & Öztin (2007), then reviewed and rephrased to meet the localized settings before distributing the survey to students. The second section contained 20 statements to determine the factors respondents found important when considering a career and how well they thought the hospitality industry offered them. On a three-point scale: "very important", "fairly important" and "not important", respondents were asked to rate each career factor in response to the question "How important is this factor to you when choosing your career?". Again on a three-point scale, "definitely", "some" and "not at all", the respondents were then asked to rate each factor in response to the question "To what extent do you think a career in tourism and hospitality will offer this factor?".

This instrument has been tested extensively by Richardson (2008, 2009a, b, 2010), where validity and reliability have been proven. In addition, a pilot sample was conducted to 5 students and 2 lecturers to check the understandability and clarity of the survey items. Respondents who filled out the questionnaire had no problems in understanding the survey questions. Hence, no changes were made to the original questionnaire.

**Data Analysis**

The data were analyzed using the Statistical Package for the Social Sciences (SPSS 18.0). Descriptive analyses (i.e., mean, and standard deviation) were employed to generate a demographic profile of the respondents and determine the importance of each factor as well as the extent to which the students believed hospitality careers offered these factors. To test statistically whether significant differences occur between the importance respondents place on career factors and the extent to which they believe hospitality offered these, a paired sample t-test was used. Ukaga and Maser (2004) stated that the paired sample t-test is used to test for differences between related or paired samples, such as when the scores or values whose means are to be compared case-for-case are from the same subject. The usual null hypothesis is that the difference in the mean values is zero. A significant difference is found if an alpha level (p-score) is less than 0.05. This test is applicable in this study as it intends to determine whether there are significant differences between the importance of factors in choosing a career and the extent to which students believe a career in tourism and hospitality will offer these factors.

- **Research Findings**

**A. Characteristics of Respondents**

This study surveyed 85 respondents (as shown in table 2), respondents tended to be male (67.1%) and (32.9%) were female. The respondents involved (25.9 %) 2nd-year students, (31.8 %) 3rd-year students, (18.8%) 4th-year students and (23.5%) postgraduates. Most of students reported that they chose to study tourism and hotel management willingly (76.5%). A sizable part of respondents (64.7%) claimed to have some experience working in the industry and approximately (32.9%) have no work experience, most of them were female.

**TABLE 2: Characteristics of the Survey Sample**

|  | Percent (%) ∗ |
|---|---|
| **Gender** | |
| • Male | 67.1% |
| • Female | 32.9% |
| **Grade** | |
| • 2$^{nd}$ year (Sophomore) | 25.9 % |
| • 3$^{rd}$ year (Junior) | 31.8 % |
| • 4$^{th}$ year (Senior) | 18.8% |
| • Postgraduate | 23.5% |
| **Study major willingness** | |
| • Yes | 76.5% |
| • No | 21.2% |
| **Practical work experience in the industry** | |
| • Yes | 64.7 % |
| • No | 32.9% |

∗ Adjusted (Valid) percentages excluding missing observations.

**B. Intention to Work in the Industry and Expectations of Future Career**

Possibly the most alarming finding to come out of this study is that, despite the unfavorable image of the industry held by students, the majority of them showed strong commitment to the industry after graduation. About (70.6%) of the students indicated that they would work in the hospitality industry after graduation while, only (5.9%) do not intend and, (23.5%) still undecided. This illustrating that the intention of students to enter the sector upon graduation is quite high. Male respondents showed a high level of intention to pursue their career in the hospitality industry where (56.6%) out of (67.1%) claimed that. Also female respondents showed the same level of intention where (28.3%) out of (32.9%) intended to work in the hospitality industry after graduation.

When asked about why they intend to pursue their career in the hospitality industry, respondents gave a list of reasons. The top reasons were "To apply the knowledge learned in in the faculty", and "Opportunities to meet and communicate with people from different cultures". Most of respondents also said that "I like this field and I want to enter after graduation to use my university degree". On the other hand, those who are not planning to work in the hospitality industry upon graduation or still undecided. Their most mentioned reasons were "low salary", "poor image about people working in tourism" and "Unsuitable work". In addition, respondents were asked an open ended question to ascertain what level of position they expected to have within three years after graduation. The majority of respondents believed that they would be in the managerial level within three years after graduation from the University and this is unrealistic in the current situation of Egypt.

**C. Factors Students Found Important When Considering a Career**

As reported in Table 3, respondents rate each item as very important, with very few respondents choosing not important for any of the factors. The most important factor identified by respondents was "A job that I will find enjoyable", where (87.1%) of respondents considered to be very important ($\bar{\chi}= 1.14$). Based on the number of respondents who chose very important as their response, the next four most important factors in choosing a career from student's view point are "A job where I gain transferable skills" ($\bar{\chi}=1.21$). "A job that is respected" approximately (78.8%) of students rate as very important ($\bar{\chi}=1.23$). "Good promotion prospects" ($\bar{\chi}=1.23$), and a job where I can use my university degree" ($\bar{\chi}=1.28$). On the other hand, only three factors received more than (10%) of respondents choosing not important including "Job mobility – easy to get a job anywhere" (10.6%), "The opportunity to travel abroad" (10.6%).

**TABLE 3: Percentages of Students' Ratings of the Importance of Factors in Choosing a Career**

| Factor | Very Important % | Fairly Important % | Not Important % | Mean $\bar{\chi}$ | Std. Deviation |
|---|---|---|---|---|---|
| 1. A job that I will find enjoyable. | 87.1 | 11.8 | 1.2 | 1.14 | .382 |
| 2. Colleagues that I can get along with. | 55.3 | 42.4 | 2.4 | 1.47 | .547 |
| 3. Pleasant working environment. | 72.9 | 24.7 | 1.2 | 1.25 | .491 |
| 4. A secure job. | 74.1 | 22.4 | 2.4 | 1.27 | .499 |
| 5. A career that provides intellectual challenge. | 64.7 | 30.6 | 2.4 | 1.47 | 1.277 |
| 6. Good promotion prospects. | 72.9 | 21.2 | 2.4 | 1.23 | .529 |
| 7. A job that gives me responsibility. | 75.3 | 21.2 | 2.4 | 1.24 | .509 |
| 8. High earnings over length of career. | 67.3 | 30.6 | 2.4 | 1.35 | .527 |
| 9. A job where I will contribute to society. | 56.5 | 35.3 | 7.1 | 1.50 | .630 |
| 10. A job where I can use my university degree. | 74.1 | 21.2 | 3.5 | 1.28 | .528 |
| 11. A job where I gain transferable skills. | 77.6 | 21.2 | 0.0 | 1.21 | .412 |
| 12. A job that is respected. | 78.8 | 15.3 | 4.7 | 1.23 | .548 |
| 13. Reasonable workload. | 50.6 | 41.2 | 4.7 | 1.50 | .612 |
| 14. A job with high-quality resources and equipment. | 68.2 | 24.7 | 7.1 | 1.38 | .619 |
| 15. The opportunity to travel abroad. | 54.1 | 35.3 | 10.6 | 1.56 | .680 |
| 16. Job mobility – easy to get a job anywhere. | 52.9 | 35.3 | 10.6 | 1.57 | .681 |
| 17. A job that can easily be combined with parenthood. | 50.6 | 40 | 8.2 | 1.55 | .663 |
| 18. Good starting salary. | 69.4 | 25.9 | 4.7 | 1.35 | .571 |
| 19. A job where I can care for others. | 57.6 | 34.1 | 8.2 | 1.50 | .647 |
| 20. A job that offers opportunities for further training. | 74.1 | 25.9 | 0.0 | 1.25 | .440 |

**Note that the mean value (1 = very important, 2 = fairly important and 3 = not important).**

## D. Perceptions and Attitudes of Hospitality Students toward a Career in the Industry

As shown in table 4, a number of statements aimed at investigating students' perceptions towards the hospitality sector. In particular, respondents were asked to express their perceptions towards 20 factors and to what extent respondents think a career in the hospitality industry offers them the factors they find important. Averaging the mean score obtained from the 20 attitudinal statements, the overall impression of the hospitality industry as a career choice was slightly favorable, with a mean score of 1.81 out of 3. Relatively speaking, more favorable perceptions were found for these items: "A job that gives me responsibility" ($\bar{\chi}= 1.54$); "Colleagues that I can get along with." ($\bar{\chi}= 1.65$). "A job where I can care for others" ($\bar{\chi}= 1.68$). At the same time, more unfavorable perceptions were found for some items where many respondents agreed that jobs in the industry does not provide opportunity to travel abroad ($\bar{\chi}= 2.05$) and does not provide good starting salary ($\bar{\chi}= 1.98$) and can't easily combined with parenthood ($\bar{\chi}= 1.97$). It is true that many tourism workers work long and unsociable hours when the rest of the population does not.

| Factor | Defiantly % | Some % | Not At all % | Mean $\bar{\chi}$ | Std. Deviation |
|---|---|---|---|---|---|
| 1. A job that I will find enjoyable. | 27.1 | 69.4 | 2.4 | 1.72 | .520 |
| 2. Colleagues that I can get along with. | 40.0 | 47.1 | 10.6 | 1.65 | .699 |
| 3. Pleasant working environment. | 38.8 | 48.2 | 11.8 | 1.70 | .687 |
| 4. A secure job. | 32.9 | 45.9 | 15.3 | 1.79 | .719 |
| 5. A career that provides intellectual challenge. | 38.8 | 44.7 | 14.1 | 1.70 | .737 |
| 6. Good promotion prospects. | 28.2 | 50.6 | 16.5 | 1.85 | .704 |
| 7. A job that gives me responsibility. | 49.4 | 41.2 | 7.1 | 1.54 | .647 |
| 8. High earnings over length of career. | 23.5 | 56.5 | 17.6 | 1.91 | .680 |
| 9. A job where I will contribute to society. | 30.6 | 42.4 | 27.1 | 1.96 | .762 |
| 10. A job where I can use my university degree. | 36.5 | 35.3 | 25.9 | 1.86 | .818 |
| 11. A job where I gain transferable skills. | 38.8 | 47.1 | 11.8 | 1.70 | .690 |
| 12. A job that is respected. | 40.0 | 49.4 | 10.6 | 1.70 | .651 |
| 13. Reasonable workload. | 30.6 | 40 | 24.7 | 1.91 | .788 |
| 14. A job with high-quality resources and equipment. | 32.9 | 55.3 | 11.8 | 1.78 | .637 |

| | | | | |
|---|---|---|---|---|
| 15. The opportunity to travel abroad. | 24.7 | 44.7 | 30.6 | 2.05 | .745 |
| 16. Job mobility – easy to get a job anywhere. | 20 | 56.5 | 22.4 | 2.02 | .658 |
| 17. A job that can easily be combined with parenthood. | 23.5 | 54.1 | 21.2 | 1.97 | .676 |
| 18. Good starting salary. | 22.4 | 52.9 | 23.5 | 1.98 | .715 |
| 19. A job where I can care for others. | 41.2 | 45.9 | 11.8 | 1.68 | .693 |
| 20. A job that offers opportunities for further training. | 35.3 | 52.9 | 10.6 | 1.72 | .661 |

**TABLE 4: Perceptions and Attitudes of Hospitality Students toward a Career in the Industry**
**Note that the Mean value 1 = definitely offers, 2 = somewhat offers and 3 = does not offer.**

Referring again to table 4, the first noticeable fact is that whilst the vast majority of respondents (more than 50%) rate these factors as very important, there are no factors where (50%) of respondents claim the industry definitely offers them. For instance, while (87.1%) of respondents claim that finding a job that is enjoyable is very important, only (27.1%) believe they will definitely find an enjoyable job in the hospitality industry. Two other attributes where major differences occur between the importance respondents place on career factors and the extent to which the hospitality industry offers the factors are related to the nature of the job. The first of these, "a secure job", where (74.1%) of respondents think this an important factor in job choice, however, only (32.9%) of respondents believed that they will find a secure job in hospitality. The second factor is "Good promotion prospects" where (72.9%) of respondents rate it as very important and only (28.2%) of them claimed that the hospitality industry will definitely offer this factor. At the same time, many students did not believe that tourism is a prestigious vocation in the society as shown by the low percentage (30.6%) of those who agreed with the statement "A job that is respected".

Another factor related to earnings asked respondents how important a "Good starting salary" was to them, with (95.3%) of respondents stating that it is important. This is in stark contrast to only (22.4%) of respondents who believe that the tourism and hospitality industry definitely offers a good starting salary, while a sizeable part of students (23.5%) claim that the hospitality industry definitely does not offer a good starting salary. The final factor that is worth discussing is "a job where I can use my university degree". The majority of respondents (74.1%) claimed this to be very important consideration when choosing a career. Although this is the case, a sizable part of students (25.9%) claimed that a job in the hospitality industry will not allow them to use their university degree and only (36.5%) of them believed that the hospitality industry will use their university degree and this is might be because most of the students believed that "it is not necessary to have a university degree to work in the tourism industry".

When testing for significant differences between the importance respondents place on career factors (table 3) and the extent to which they believe hospitality sector offered these (table 4), every factor is seen to be significantly different as their p-value (0.00) is less than the critical value of 0.05. In each of these factors the importance factor has a lower mean than the extent to which students believe a career in tourism and hospitality offers that factor. This infers that students do not believe that a career in tourism and hospitality will offer them the factors that they find important in choosing a future career. These results correspond closely to the findings of Richardson & Butler (2011), who explored Malaysian undergraduate tourism and hospitality students' views of the industry as a career choice and Richardson's (2009b), who found that students in Australia have similar feelings about the factors that the industry offers to potential employees.

### E. Variables Affecting Students' Perceptions Of and Attitudes toward the Hospitality Industry as A career Choice.

This part measures the relationship between several variables that may have effects on hospitality students' perceptions and attitudes toward the industry and their career in the industry. These contingency variables are: (1) gender, (2) work experience, (3) years of study, and (4) study major willingness.

**Gender**

An independent-sample t-test was conducted to examine if significant differences were found between respondents perceptions of hospitality careers according to gender. The results show that female students tend to have less favorable perceptions ($\bar{\chi}=1.92$) than male students ($\bar{\chi}=1.76$) but, generally there was no significant gender-based difference in the perception of students toward the hospitality careers (see table 5).

**Work Experience**

Tests were also conducted to examine if there were statistically significant correlations between the respondents having (or not having) relevant work experience and their perceptions of the industry as a career choice. Students who have practical work experience in the industry tend to have a more positive attitude ($\bar{\chi}=1.78$), than those who did not work for hotels before ($\bar{\chi}=1.85$).

**Study Major Willingness**

Referring again to table 5, the test indicated that the mean value of the perceptions of respondents who were willing to study tourism ($\bar{\chi}=1.76$) is much more favorable than those who were not willing ($\bar{\chi}=2.01$). A significant positive correlation was found between student's willingness to study tourism and their perceptions towards the industry as a career choice.

**Year of Study**

Finally, statistically significant correlations between years of study and career perceptions of the respondents were found. Post-hoc comparisons made using the Tukey HSD test showed senior students are more likely to have positive attitudes towards the industry ($\bar{\chi}=1.52$), followed by junior students ($\bar{\chi}=1.80$) then sophomores ($\bar{\chi}=1.85$). In other words, as the respondents progress in their degree, their perceptions of the industry improved. On the other hand, postgraduates had the least favorable image towards the industry ($\bar{\chi}=2.01$).

**Table 5: Variables Affecting Students' Perceptions of and Attitudes toward the Hospitality Industry as A career Choice.**

| Variables | $\bar{\chi}$ | SD | T | Sig. (2tailed) |
|---|---|---|---|---|
| Gender | | | | |
| • Male | 1.76 | .340 | 1.988 | 0.050 |
| • Female | 1.92 | .351 | | |
| Work experience | | | | |
| • Yes | 1.78 | .366 | -.819 | 0.415 |
| • No | 1.85 | .332 | | |
| Study major willingness | | | | |
| • Yes | 1.76 | .346 | | |
| • No | 2.01 | .338 | -2.687 | 0.009 |
| Year of Study (Grade) | | | | |
| • Sophomore | 1.85 | .272 | | |
| • Junior | 1.80 | .331 | | |
| • Senior | 1.52 | .262 | | |
| • Postgraduate | 2.01 | .387 | F= 7.184 | 0.000 |

**Note: n=85**

**Discussion and Implications**

This research was undertaken to find out whether hospitality students intend to pursue their career in the industry and how well they thought the hospitality sector will offer them the desired career. Based on the study findings, the majority of respondents have a strong intention to enter the hospitality industry after graduation which contrasts with the conclusions of some previous studies reported by Jenkins (2001) and Richardson (2008) but concurs with the findings of Roney and Öztin (2007). Although students show high intention to enter the industry upon graduation, they do not believe that the industry will offer them the career they seek. This suggests that hotels may be able to attract new graduates to join the workforce because of the lack of job choice in the market. But, when students find a job offer in other better paying industries they will pursue career outside tourism, which will negatively impact on the quality and quantity of future staff, the government that has invested money in tourism education and the students whom spent years studying tourism courses.

Students generally have a neutral attitude towards jobs in the industry however, the results reveal that respondents perceived hospitality jobs as stressful with unreasonable workloads, which echoes the findings of some previous studies of hospitality settings (Kusluvan & Kusluvan, 2000; Richardson, 2008). In terms of earnings and salary, hospitality jobs are perceived as low salary opportunities especially in the begging of the career path. This may be one of the main reasons behind intending to work in other better paying industries after graduation. Students also have unrealistic perceptions related to the position they expected to have, where most of them claimed to be in a managerial position within three years after graduation. This agrees with Casado (1992) and Sciarini (1997) who found that most graduates believed they were qualified enough to work as an assistant manager and were looking for a position that was of a managerial level rather than an hourly operational position. Hence, it is good for the students to recognize the reality of future working conditions because they will form more realistic and lower expectations with regard to jobs in the industry.

Also, few respondents believed that working in hospitality provide a secure job and this reflects their impressions about the current situation in Egypt where a sizable part of hotel work force suffered from discharging or unpaid vacations during recent crises. This is in addition to the nature of employment in the tourism industry which is notoriously insecure because of seasonality, fluctuations in demand, and the high number of part-time and temporary jobs (Bull, 1995). Students also believe that any one can work in tourism even those with lower academic qualifications which concurs with the results of Roney & Öztin, (2007). Therefore, hospitality professors should strive to reverse this negative perception towards tourism industry, by projecting their true qualities and uniqueness of such an educational experience.

As students' progress in their studies and have work experience in the industry, their perception of hotel-related jobs is affected in a positive way. This contrasts with the findings of many previous studies, such as Barron and Maxwell (1993), Jenkins (2001), and Richardson (2008), but, highlights the importance of the design and implementation of effective internship programs. So that educators should continue to refine internship programs that promote students' professional growth as well as realistic expectations of the industry. Those who were willing to study tourism as a first priority, show much more positive attitudes toward the industry than others. So that new students should be carefully interviewed and selected based on their willingness to study tourism as one of the main requirements.

Finally, one of the issues facing employers is to understand the new generation of employees entering the workforce. Although respondents rate each item as very important, the top factors students revealed to be very important to them when selecting a career are; "A job that is enjoyable" "A job where I gain transferable skills", "A job that is respected", "Good promotion prospects" and a job where I can use my university degree". At the same time students who are currently studying hotel management do not believe that working in hospitality related jobs will offer them these factors which corresponds closely to the findings of Richardson & Butler (2011), who explored Malaysian undergraduate tourism and hospitality students' views of the industry as a career choice and Richardson's (2009b), who found that students in Australia have similar feelings about the factors that the industry offers to potential employees. This highlights the fact that unless the industry can meet the expectations of students, it will continue to lose those highly skilled and trained potential employees.

Another important finding to be discussed is that, while most of respondents claimed "a job where they can use their university degree" is a very important factor when selecting a career, it unfortunately, becomes a major problem when fresh graduates who want to apply their academic knowledge, start their career with old-mental managers who told them "ignore or forget what you have studied in the faculty". This situation explains the results of Zou *et al.* (2002), who concluded that about 50% of the graduates who entered the tourism industry upon graduation quit their first job and found employment outside the industry after 2 years of working. This results in high staff turnover and waste of trained and experienced personnel. Hence, industry professionals and employers should deal carefully with recent recruits from students who have a major effect on the potential progress of the industry.

**Conclusion and Recommendations**

The current study investigated student's attitudes towards the hospitality industry as career choice. This study shows that although the majority of students intend to work in the hospitality after graduation, they still do not see the industry as an appealing career path because many of the factors they find important in a career are missing. These findings highlight the need for the industry to adopt tactics and strategies aimed at ensuring that potential employees, i.e. tourism and hospitality students, are not leaving the industry or even failing to enter the industry on graduation. One of the research objectives was to recommend a set of specific remedial actions that could be initiated by hospitality stakeholders to improve the image of the industry as a career choice. The following part provides recommendations directed for (a) The Government (b) The Industry Employers and (c) University Leaders and Educators.

**A. The Government**

- **Government Should Fill The Gap Between The Industry and University.** It's fundamental for the government that invests a lot of money on tourism education to take some measures to ensure that hospitality graduates enter the industry upon graduation. Hence, the government, and especially the Ministry of Tourism needs to fill the gap (the period of time from student graduation till starting his\her career in hospitality related jobs), during this period students may fail to enter the industry and this ends in completing their career in other industries. This can be conducted by; establishing a unit that aims at linking fresh graduates to hospitality recruiters. So that, graduated students could be distributed on the top hotels to fill the vacant jobs every year.

- **Establishment of the Hospitality Managers and Educators Collaborative Forum.** This forum will strive to facilitate a long-term mutually beneficial relationship between industry professionals and educators. It will benefit immeasurably the planning and implementation of internship practices by highlighting deficiencies and indicating ways to bridge the gap between educational theory and actual practice.

- **General Tourism Awareness Activities.** This could take a number of shapes and forms ranging from informational campaigns, organization of presentations by prominent industry leaders, permissions to conduct educational field trips to hospitality establishments, participation in training programs for students and graduates.

### B. The Industry Employers

- Tourism and hospitality organizations must continue to work on improving many aspects of the working conditions within the industry. There are a number of areas that students believe the industry is behind other industries, particularly in regards to pay, promotion opportunities, career prospects and job security. Unless the industry can change the perceptions of a career in the industry it will continue to lose these highly skilled and trained employees.

- Industry organizations are also encouraged actively to seek partnership with the university to design internship programs for the hospitality students that ensure the students have positive experiences. This creates a win–win situation in which both parties can benefit.

- **Word of mouth from graduated students.** The high degree of importance that word of mouth from students who started their career in hospitality plays in developing student perceptions towards the industry confirms that there is likely a spillover effect that occurs among undergraduate students. That's why; the hospitality managers should deal carefully with newcomers who have a major effect on potential recruits perceptions of the industry.

### C. University Leaders and Educators

- **Organized Tours to Local Hospitality Establishments**. The faculty council in collaboration with the local industry professionals should organize continual field trips to prominent hospitality establishments in which students will have the opportunity to experience the true qualities of the industry.

- **Educators need to play their part** in ensuring that students are being given realistic expectations of the types of position available in the industry, pay levels, promotion opportunities, and career paths. They must work more closely with industry partners when designing future curricula.

- **University students are internet savvy**. With the necessary planning and preparation internet and social media could be utilized to provide information and generate interest towards the industry. The faculty should encourage the development of an innovative and interactive internet web site that directly targets tourism students and graduates. The primary objective of this site is to provide a mean of communication between students, educators and industry professionals.

- **Improve the Image and Project the Genuine Qualities of Tourism Education**; unfortunately, a sizable part of the Egyptian society perceives a poor image about working in tourism and hospitality professions. As they believe that any one can work in tourism even those with lower academic qualifications. Hospitality professors should strive to reverse this negative perception towards tourism industry, by projecting their true qualities and uniqueness of such an educational experience.

## Research Limitations and Further Research

The findings of this study are useful; but it is not without limitations. Only hospitality students from Alexandria University were surveyed, which means that it is difficult to generalize from the data analyzed. Further research would consider extending the number of participating faculties and institutions. This approach would enhance both the credibility and generalizability of the topic under investigation. Additionally, this research adopts a quantitative approach. Further research in the same area could be done by undertaking qualitative assessments, which will provide more in-depth understandings of the students' perceptions and attitudes. Further study could also use the same questionnaire to see if the same effects and attitude apply to students in the coming years.

## Acknowledgment

The researcher is grateful to Dr. M. Shabaan and Miss. D. Abd-Elhamid for their valuable comments on the earlier draft of the article. Considerable thanks also extended to Dr. I. Salem and Miss. S. Samir for helping with the statistical analysis.